\documentclass[preprint,twocolumn,5p,authoryear,colorlinks=true]{elsarticle}

\usepackage{lineno,hyperref}
\usepackage{float,dblfloatfix}

\journal{Icarus}

\usepackage[labelfont=bf,justification=justified,singlelinecheck=false]{caption}
\usepackage{numcompress}

\begin{document}

\begin{frontmatter}
\title{Aerosols optical properties in Titan's Detached Haze Layer before the equinox}

\author[URCA]{Beno\^{i}t Seignovert\corref{correspondingauthor}}\ead{research@seignovert.fr}
\author[URCA]{Pascal Rannou}
\author[URCA]{Panayotis Lavvas}
\author[URCA]{Thibaud Cours}
\author[JPL,URCA]{Robert A. West\fnref{visiting}}

\address[URCA]{GSMA, Universit\'{e} de Reims Champagne-Ardenne, UMR 7331-GSMA, 51687 Reims, France}
\address[JPL]{Jet Propulsion Laboratory, California Institute of Technology, Pasadena, CA 91109, USA}

\cortext[correspondingauthor]{Corresponding author}
\fntext[visiting]{Visiting at GSMA/URCA}

\begin{abstract}
UV observations with Cassini ISS Narrow Angle Camera of Titan's detached haze is an excellent tool to probe its aerosols content without being affected by the gas or the multiple scattering. Unfortunately, its low extent in altitude requires a high resolution calibration and limits the number of images available in the Cassini dataset. However, we show that it is possible to extract on each profile the local maximum of intensity of this layer and confirm its stability at $500 \pm 8$~km during the 2005-2007 period for all latitudes lower than 45$^\circ$N. Using the fractal aggregate scattering model of \cite{Tomasko2008} and a single scattering radiative transfer model, it is possible to derive the optical properties required to explain the observations made at different phase angles. Our results indicates that the aerosols have at least ten monomers of 60~nm radius, while the typical tangential column number density is about $2\cdot 10^{10}$~agg.m$^{-2}$. Moreover, we demonstrate that these properties are constant within the error bars in the southern hemisphere of Titan over the observed time period. In the northern hemisphere, the size of the aerosols tend to decrease relatively to the southern hemisphere and are associated with a higher tangential opacity. However, the lower number of observations available in this region due to the orbital constraints is a limiting factor in the accuracy of these results. Assuming a fixed homogeneous content we notice that the tangential opacity can fluctuate up to a factor 3 among the observations at the equator. These variations could be linked with short scale temporal and/or longitudinal events changing the local density of the layer.
\end{abstract}

\begin{keyword}
Titan \sep Atmosphere \sep Aerosols

doi:\href{https://dx.doi.org/10.1016/j.icarus.2017.03.026}{10.1016/j.icarus.2017.03.026}
\end{keyword}
\end{frontmatter}


\section{Introduction}
Since the beginning of exploration of giant planets in the 80's, Titan, the largest moon of Saturn, always focused a lot of attention on its unique thick haze atmosphere. Voyager 1 and 2 flybys, revealed the existence of a complex stratified succession of haze layers above Titan's main haze layer \citep{Smith1981,Smith1982}. One of these layers, the Detached Haze Layer (DHL), presented a large horizontal extent at 350~km and could be seen all along the limb surrounding Titan main haze between 90$^\circ$S up to 45$^\circ$N before merging with the north polar hood. Based on Voyager 2 radial intensity scans at high phase angles, \cite{Rages1983b} were able to retrieve the vertical distribution of scattering particles and reveal an important depletion of the aerosol particle density below this layer. \cite{Toon1992} proposed the first explanation by an interaction of ascending winds with the vertical structure of the haze, yielding a secondary layer above the main layer. The aerosols are initially produced in a single production zone in the main haze then raised by the dynamics and stored inside the detached haze layer. On the other hand, \cite{Chassefiere1995} presented a completely different formation scenario based on the photochemistry of polyacetylenes occurring at high altitudes producing fluffy aggregates trapped inside the detached haze layer. Finally \cite{Rannou2002,Rannou2004} proposed with a global circulation model (GCM) that the detached haze is an natural outcome of the interaction between the atmosphere dynamics and the haze microphysics. Therefore, the detached haze is submitted to a seasonal cycle driven by dynamics.

The arrival of the Cassini spacecraft in the Saturnian System in 2004, was a unique opportunity to investigate the persistence of the detached haze layer over time. \cite{Porco2005} confirmed its presence 24 years after its discovery at an altitude near 500~km. This new location was first explained by \cite{Lavvas2009} as a secondary layer under the limit of the Voyager camera sensitivity. Their mechanism based on the sedimentation and coagulation of particles of mono-dispersed spheres was able to produce a detached haze at 500~km altitude independently of the dynamical transport. However, the systematic survey made by the repetitive flybys of Titan by Cassini showed that its altitude at the equator remains constant at 500 km between 2005 and 2007, decreases to 480~km at the end of 2007 and dropped suddenly at 380~km in 2009, around the northern spring equinox, before disappearing in 2012 \citep{West2011}. Finally, recent 3D GCMs \citep{Lebonnois2012,Larson2015}, more sophisticated than the previous 2D GCM, still producing the detached haze, give a description of the complete annual cycle of the detached haze layer and predict a reappearance in early 2015.

Our purpose is to characterize the physical properties of the detached haze layer before the equinox, while it was stable. We analyzed ISS images of the detached haze layer at different phase angles at UV wavelengths where the detached layer can be easily seen. Then we interpret the observations using a single-scattering radiative transfer model \citep{Tomasko2008} with fractal aggregate particles to constrain the aerosol optical properties based on the monomer radius, the number of monomers per aggregate and tangential column number density. Finally we derive the local temporal/longitudinal variability of the detached layer during the 2005-2007 period.

\section{Observations}
\begin{figure}[ht!]\centering
\includegraphics[width=.5\textwidth]{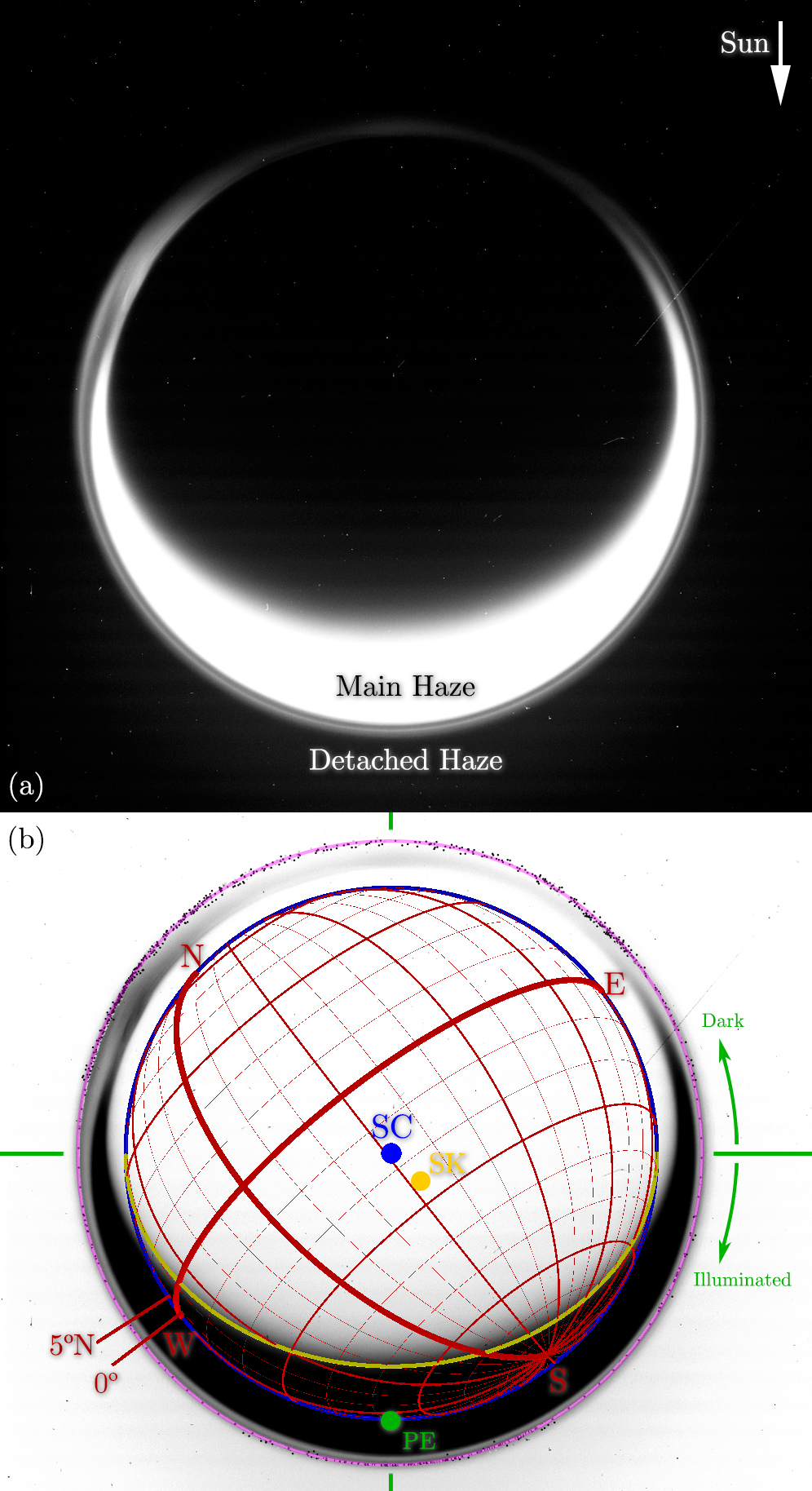}
\caption{\label{fig:iss_georef}(a) Overexposed and contrast-enhanced ISS NAC N1551888681\textunderscore 1 image taken on 2007/03/06 at an observation phase angle of 143$^\circ$ in CL1-UV3 filters (338~nm). The detached haze layer can be seen on the limb surrounding Titan's main haze. (b) Same image inverted. The small black dots represents the detected position of local maxima associated to the detached haze layer fitted by an ellipse (magenta). The center of the fit provides the location of Titan's center (i.e. the sub-spacecraft point (SC) in blue at 31$^\circ$E - 17$^\circ$S). The orange point is given as a reference for the Spice Kernels (SK) predicted center. An offset of 55 pixels (430~km) can be noticed. The red lines are the geographic coordinates. The sub-solar point is located on the other side of Titan at 127$^\circ$W - 13$^\circ$S and the yellow line represents the location of the terminator on the ground. The green cross behind Titan represents the photometric frame with a photometric equator (PE) at 79$^\circ$W - 49$^\circ$S. The limb profiles are average in $I/F$ in 5$^\circ$ bins in latitude on the illuminated side of Titan.}
\end{figure}

To determine the properties of the aerosols in the detached haze layer, we made a survey over the ISS images taken with the Narrow Angle Camera (NAC) before its first decrease in altitude at the end of 2007 \citep{West2011}. We limit our analysis to the CL1-UV3 filter ($\lambda$=338~nm) to get the best contrast between the detached haze layer and the main haze (Fig.~\ref{fig:iss_georef}a) and to minimize the multiple scattering. Assuming lambertian scattering and the albedo observations by~\cite{Karkoschka1994,Karkoschka1998}, we estimate that the multiple scattering is always less than 5\% for UV wavelengths. We select images taken far enough from Titan to get the best latitude coverage on the limb and to get an accurate georeference calibration of data, but also close enough to get the highest pixel scale to probe properly the detached haze layer ($<$10~km). We also restrict our analysis to a short period of time between 2005 to 2007 to limit the temporal variability of the detached haze layer (1 Titan year = 29 Earth years). Among the acquired images only 15 (Tab.~\ref{tab:1}) satisfied the above criteria, covering a large range of phase angle (from 14$^\circ$ to 155$^\circ$).

\begin{table}[ht!]
\centering
\caption{\label{tab:1}Dataset of 15 ISS NAC images used in this study, taken with CL1-UV3 filter combination (338~nm). The phase angle, the limb latitude coverage, the latitude of the photometric equator and the pixel scale are also listed below.}
\scriptsize
\begin{tabular}{l l l l p{.5cm} p{.5cm} }
\hline
ISS ID                        & Date        & Phase      & Coverage                 & Lat. Photo. & Pixel scale (km) \\
\hline
N1506288442\textunderscore 1	& 2005/09/24	& 145$^\circ$ & 90$^\circ$S-25$^\circ$N & 37$^\circ$S & 5.5 \\
N1506300441\textunderscore 1	& 2005/09/25	& 151$^\circ$ & 90$^\circ$S-30$^\circ$N & 46$^\circ$S & 5.5 \\
N1509304398\textunderscore 1	& 2005/10/29	& 155$^\circ$ & 90$^\circ$S-25$^\circ$N & 51$^\circ$S & 6.1 \\
N1521213736\textunderscore 1	& 2006/03/16	&  68$^\circ$ & 90$^\circ$S-65$^\circ$N & 19$^\circ$S & 7.3 \\
N1525327324\textunderscore 1	& 2006/05/03	& 147$^\circ$ & 90$^\circ$S-55$^\circ$N & 33$^\circ$S & 7.2 \\
N1540314950\textunderscore 1	& 2006/10/23	& 120$^\circ$ & 65$^\circ$S-55$^\circ$N &  2$^\circ$S & 5.7 \\
N1546223487\textunderscore 1	& 2006/12/31	&  66$^\circ$ & 70$^\circ$S-65$^\circ$N &  6$^\circ$S & 7.4 \\
N1551888681\textunderscore 1	& 2007/03/06	& 143$^\circ$ & 75$^\circ$S-35$^\circ$N & 49$^\circ$S & 7.9 \\
N1557908615\textunderscore 1	& 2007/05/15	&  25$^\circ$ & 80$^\circ$S-10$^\circ$N & 76$^\circ$S & 7.8 \\
N1557919415\textunderscore 1	& 2007/05/15	&  25$^\circ$ & 80$^\circ$S-10$^\circ$N & 75$^\circ$S & 8.2 \\
N1559282756\textunderscore 1	& 2007/05/31	&  20$^\circ$ & 85$^\circ$S-10$^\circ$N & 80$^\circ$S & 7.7 \\
N1562037403\textunderscore 1	& 2007/07/02	&  14$^\circ$ & 90$^\circ$S-30$^\circ$N & 58$^\circ$S & 7.8 \\
N1567440117\textunderscore 1	& 2007/09/02	&  20$^\circ$ & 90$^\circ$S-55$^\circ$N & 22$^\circ$S & 7.9 \\
N1570185840\textunderscore 1	& 2007/10/04	&  27$^\circ$ & 90$^\circ$S-60$^\circ$N & 28$^\circ$S & 7.3 \\
N1571476343\textunderscore 1	& 2007/10/19	&  80$^\circ$ & 85$^\circ$S-60$^\circ$N & 11$^\circ$S & 8.2 \\
\hline
\end{tabular}
\end{table}

The radiance factor on the ISS images ($I/F$) is calibrated using the CISSCAL routines \citep{West2010} and a Poisson Maximum-a-posteriori (PMAP) deconvolution is applied to improve the signal/noise ratio. Then the detached haze layer is automatically located at the maximum of contrast. Considered to be stable during this period \citep{West2011}, we use its altitude as a proxy to locate very precisely the center of Titan (Fig.~\ref{fig:iss_georef}b) by fitting an ellipse through these points. Therefore we are able to extract the $I/F$ profile all along the limb of Titan according to the geographical coordinates and the illumination conditions during the acquisition.

\begin{figure*}[hb!]\centering
\includegraphics[width=\textwidth]{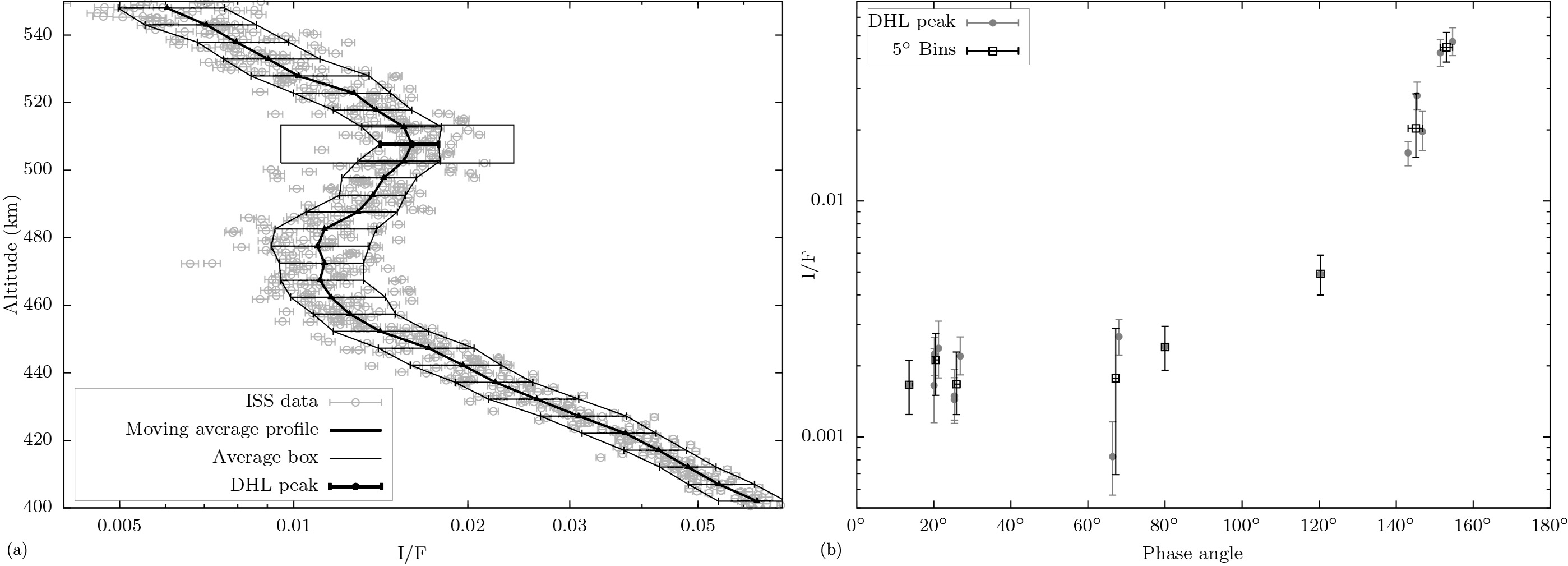}
\caption{\label{fig:data}(a) Intensity $I/F$ vertical profile measured from ISS NAC in CL1-UV3 filter combination on N1551888681\textunderscore 1 close to the geographic equator of Titan (0 to 5$^\circ$N at -55$^\circ$ from the photometric equator). Each pixel is considered with its own photon noise uncertainty (grey circles). The profile is smoothed with a moving average box (dashed black box) of 1.5 times the pixel scale of the image (e.g. 10.4 km here). The intensity spread of the profile (black lines) corresponds to the normalized Gaussian cumulative uncertainty at 1$\sigma$. (b) Collection of local maxima $I/F$ of the detached haze layer (dots) close to the geographic equator of Titan (0 to 5$^\circ$N) over the different phase angles. Data used in the model (black squares) were integrated with a 5$^\circ$ bin using Gaussian uncertainty integration with a 1$\sigma$ Gaussian equivalent to keep the spatial and temporal spread of the data. For each latitude only the local maximum $I/F$ on the detached haze layer and its spread is used in the rest of the study.}
\end{figure*}

The vertical $I/F$ profile is sampled in latitude using 5$^\circ$ bins (Fig.~\ref{fig:data}a). The uncertainty of the intensity on each pixel represents the photon noise (between 2 to 5\%) and the flat field uncertainty (1\%). To account for the observed variability of the profile inside the latitude bin and altitude ranges, we sum-up all pixel uncertainties with a moving average box of 1.5 times the pixel scale (i.e with an overlap of 50\%) and containing at least 30 pixels (up to 150).  Assuming a stochastic uncertainty distribution, the sum of these uncertainties is considered Gaussian and the value at 0.16 and 0.84 on the normalized cumulative uncertainty are equivalent to a Gaussian 1$\sigma$ error (i.e. covering 68\% of the integral). This uncertainty distribution provides average and variance values of the spatial variability. This method allows us to smooth the vertical profile and remove the outliers. Finally the local maximum $I/F$ of the detached haze layer is extracted for each profile at all latitudes in the illuminated limb of Titan.

This process is applied to the whole image dataset. Figure~\ref{fig:data}b represents the distribution of the local maximum $I/F$ of the detached haze layer as a function of phase angle. The variability of the local maximum $I/F$ of the detached haze layer is resampled using bins of 5$^\circ$ phase angle range in order to put similar weight on the different images at low and high phase angles.

\section{Method}
As mention before, we focus our study on the local maximum $I/F$ of the detached haze layer. We consider that above the detached haze layer, the atmosphere of Titan is optically thin and the incoming flux from the Sun is not significantly attenuated down to the detached haze layer (i.e. the opacity along the incoming ray $\tau^0_\mathrm{ext}\ll 1$ in the illuminated limb). We also neglect the multiple scattering inside and below the detached haze layer (evaluated at maximum to 5\% at 338~nm). Then, in the limit of the single scattering approximation and the optically thin layers, the output flux can be simply described as:

\begin{equation}
I/F =  \frac{ \omega \cdot P(\theta)}{4} \cdot
\left[ 1 - \exp\left(-\mathrm{N}_\mathrm{los} \cdot
\sigma_\mathrm{ext} \right)\right]
\end{equation}

with $\omega$ the single scattering albedo, $P(\theta)$ the phase function at the scattering angle $\theta$ (corresponding to $180^\circ$ - phase angle of the observation), $\sigma_\mathrm{ext}$ the extinction cross-section of the aerosols, and $\mathrm{N}_\mathrm{los}$ the tangential column number density (i.e. the amount of aggregates along the line of sight).

We assume that the detached haze layer is composed of fractal aerosols composed of single-size spherical primary particles called monomers \citep{West1991,Cabane1993}. Based on a tholin composition, we use the optical index n=1.64 and k=0.17 \citep{Khare1984} for the ISS CL1-UV3 filter ($\lambda$=338~nm). Then, assuming a fractal dimension (D$_f$) of 2.0 and a maximum number of monomers per aggregate $N \le 5,000$ \citep{Tomasko2009}, the optical properties ($\omega$, $P(\theta)$ and $\sigma_\mathrm{ext}$) of theses aerosols can be calculated using the model developed by \cite{Tomasko2008}. Therefore, the aggregate itself can be described by only two parameters: R$_m$ the monomer radius and $N$ the number monomers per aggregate. The bulk radius (R$_v$) is defined with the relation:

\begin{equation}
\mathrm{R}_v = \sqrt[3]{N} \cdot \mathrm{R}_m
\end{equation}

Between 2005 and 2007, we assume that the detached haze is stable and the aerosol content of the detached haze layer, i.e. the number of aggregates and their optical properties remain temporally constant~\citep{West2011} and also longitudinally stable due to mixing by zonal winds \citep{Larson2015}. Then we calculate a synthetic $I/F_\mathrm{synt}$ with a set of 3 independent parameters: R$_m$, $N$ and $\mathrm{N}_\mathrm{los}$, that we compare with the observed $I/F_\mathrm{obs}$ as a function of phase angle for the different latitudes. This comparison is made with a $\chi^2$ as a merit function~\citep{Press1992}:

\begin{equation}
\chi^2(\mathrm{lat}) = \sum\limits_{\phi} \left(
\frac{I/F_\mathrm{obs}^{\phi,\mathrm{lat}} - I/F_\mathrm{synt}^{\phi,\mathrm{lat}}(\mathrm{R}_m,N,\mathrm{N}_\mathrm{los})}
{ \Delta I/F_\mathrm{obs}^{\phi,\mathrm{lat}} }
\right) ^2
\end{equation}

with $\phi$ the phase angle, and $\Delta I/F_\mathrm{obs}^{\phi,\mathrm{lat}}$ the uncertainty associated to observation $I/F_\mathrm{obs}^{\phi,\mathrm{lat}}$.

In practice, the mapping of the $\chi^2$ is performed with a 2~nm step between 10 and 70~nm for the monomer radius (R$_m$), and 10~points per decade between 1 and 5,000 for the number of monomers per aggregates ($N$). The upper limit on $N \le 5,000$ correspond to typical aggregates observed close to the surface \citep{Tomasko2009}. We use a mono-dispersed distribution of particles to increase the speed of the calculations. The tangential opacity ($\tau_\mathrm{ext} = \mathrm{N}_\mathrm{los} \times \sigma_\mathrm{ext} $) is directly linked to the scaling factor between the phase function and the intensity and is fitted with a least squares minimization for all pairs of (R$_m$, $N$).

\section{Results}
\begin{figure}[ht!]\centering
\includegraphics[width=.5\textwidth]{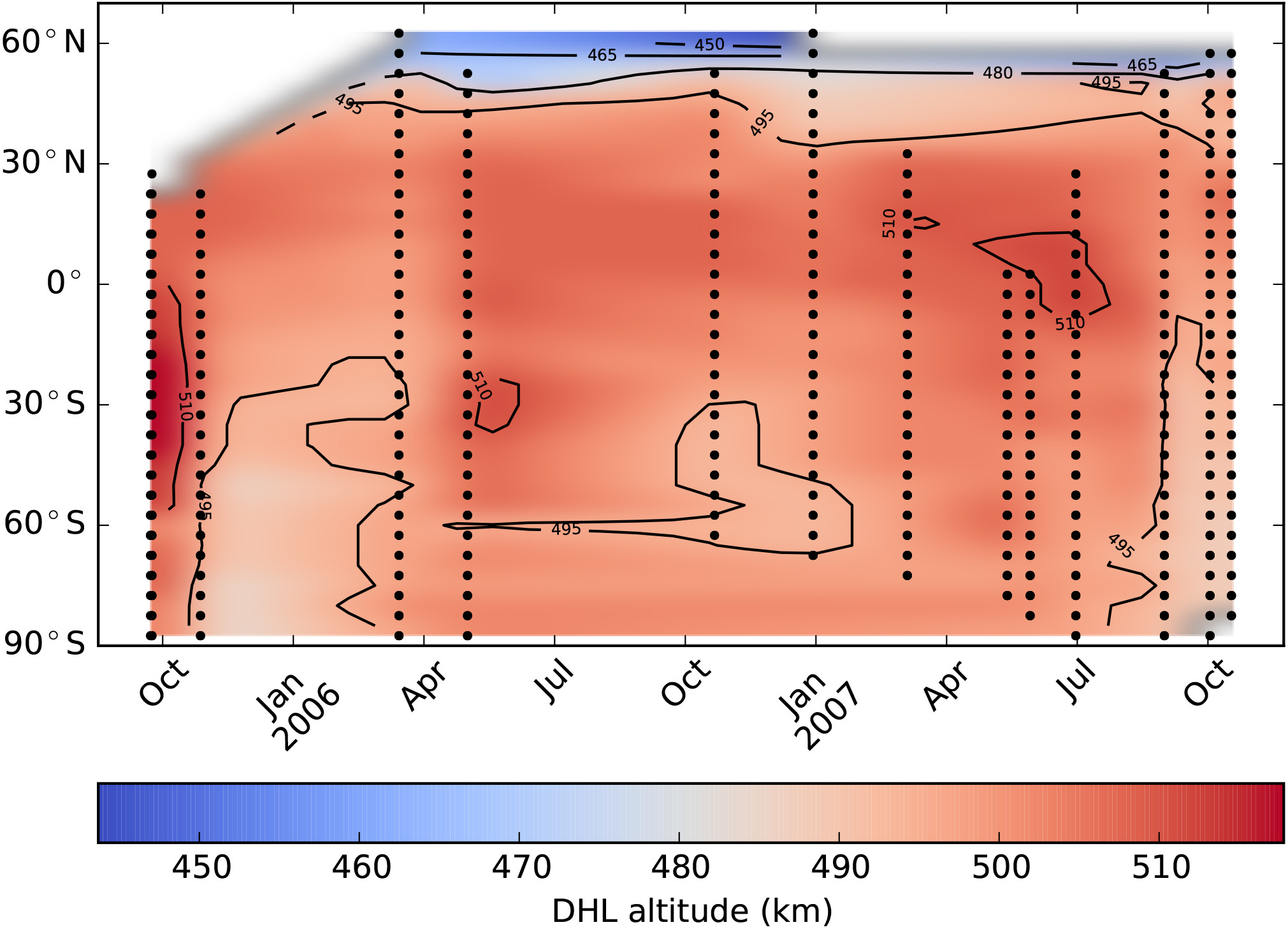}
\caption{\label{fig:alt}Altitude map at the local maxima $I/F$ on the detached haze layer as a function of the time and latitude. The black dots represent the profiles analyzed.}
\end{figure}

The altitude of the local maximum $I/F$ of the detached haze layer for the complete dataset (Fig.~\ref{fig:alt}) presents a high stability at $500 \pm 8$~km. Thus we confirm that the detached haze layer is a persistent feature, not only at the equator \citep{West2011}, but at all latitudes below 45$^\circ$N during the 2005-2007 period. The values for the latitudes higher than 45$^\circ$N are affected by the polar hood observed at this period \citep{Griffith2006}.

\begin{figure*}[ht!]\centering
\includegraphics[width=\textwidth]{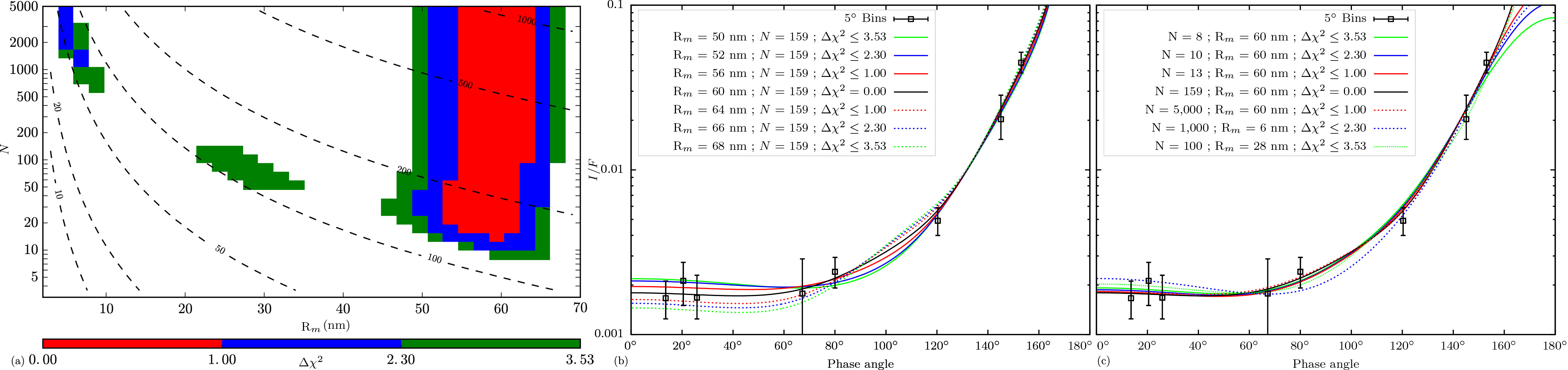}
\caption{\label{fig:chi2}(a) (R$_m$, $N$) pairs explored for latitudes close to Titan's geographic equator (0 to 5$^\circ$N). The confidence limits $\Delta\chi^2 = \chi^2 - \chi^2_\mathrm{min}$ are color coded with respect to $\chi^2_\mathrm{min}$=1.7 at R$_m$=60~nm and $N$=125. The black dash lines represent the aggregate bulk radius (R$_v$). (b) Evolution of $I/F_\mathrm{synt}$ for different R$_m$ with $N$ fixed at 125 monomers per aggregate. The main changes are observed in the backscattering direction. (c) Same outputs but for different $N$ with R$_m$ fixed at 60~nm. In this case, only the forward peak is changed. Unfortunately, observations at very high phase angles were not available and do not provide an upper limit on $N$ (e.g. $N=159$ and $N=5,000$ are identical up to 170$^\circ$). We also present on the same plot samples of the two local minima observed in (a). These parameters fit some observations but are not consistent with the others.}
\end{figure*}

For each latitude bin, we compute the synthetic $I/F$ for all pairs of (R$_m$, $N$) and we keep the best fit on the tangential opacity. All the solutions explored in the bin close to the geographic equator of Titan (0 to 5$^\circ$N) are represented on Figure~\ref{fig:chi2}a. Based on the \cite{Press1992} description of the confidence limits on estimated model parameters (chapter 15.6), we color code the $\Delta\chi^2 = \chi^2 - \chi^2_\mathrm{min}$ for a confidence level $p=68.3\%$. Thus all fits with $\Delta\chi^2 <$ 3.53 (i.e. with 3 degree of freedom) match the observations within their error bars (Fig.~\ref{fig:chi2}b and 4c). For this latitude (0 to 5$^\circ$N), the $\chi^2_\mathrm{min}$ is achieved at R$_m = 60$~nm, $N = 159$ and N$_\mathrm{los} = 3.7 \cdot 10^{10}$~agg.m$^{-2}$ with a value of 1.7. Then we consider each parameter independently (i.e. with 1 degree of freedom), so we select all the pairs of (R$_m$, $N$) for which $\Delta\chi^2 <$ 1.00. At this latitude minima and maxima obtained are R$_{m,\mathrm{min}} = 54$~nm, R$_{m,\mathrm{max}} = 64$~nm, $N_\mathrm{min} = 13$ and $N_\mathrm{max} = 5,000$ (upper limit tested). For these parameters, the size of the monomers (R$_m$) is constrained by the backscattering part of the phase function with the observations at low phase angle (Figure~\ref{fig:chi2}b). On the other hand, the determination of the number of monomers per aggregate ($N$) is mainly driven by the shape of the forward peak (Figure~\ref{fig:chi2}c). Unfortunately, observations at very high phase angles ($>160^\circ$) were not available in the CL1-UV3 filter during this period, and do not allow us to constrain the upper limit of $N$. Then, the bulk size of the aggregate (R$_v$) is poorly constrained between 150~nm to 1.1~$\mu$m.

\begin{figure*}[ht!]\centering
\includegraphics[width=\textwidth]{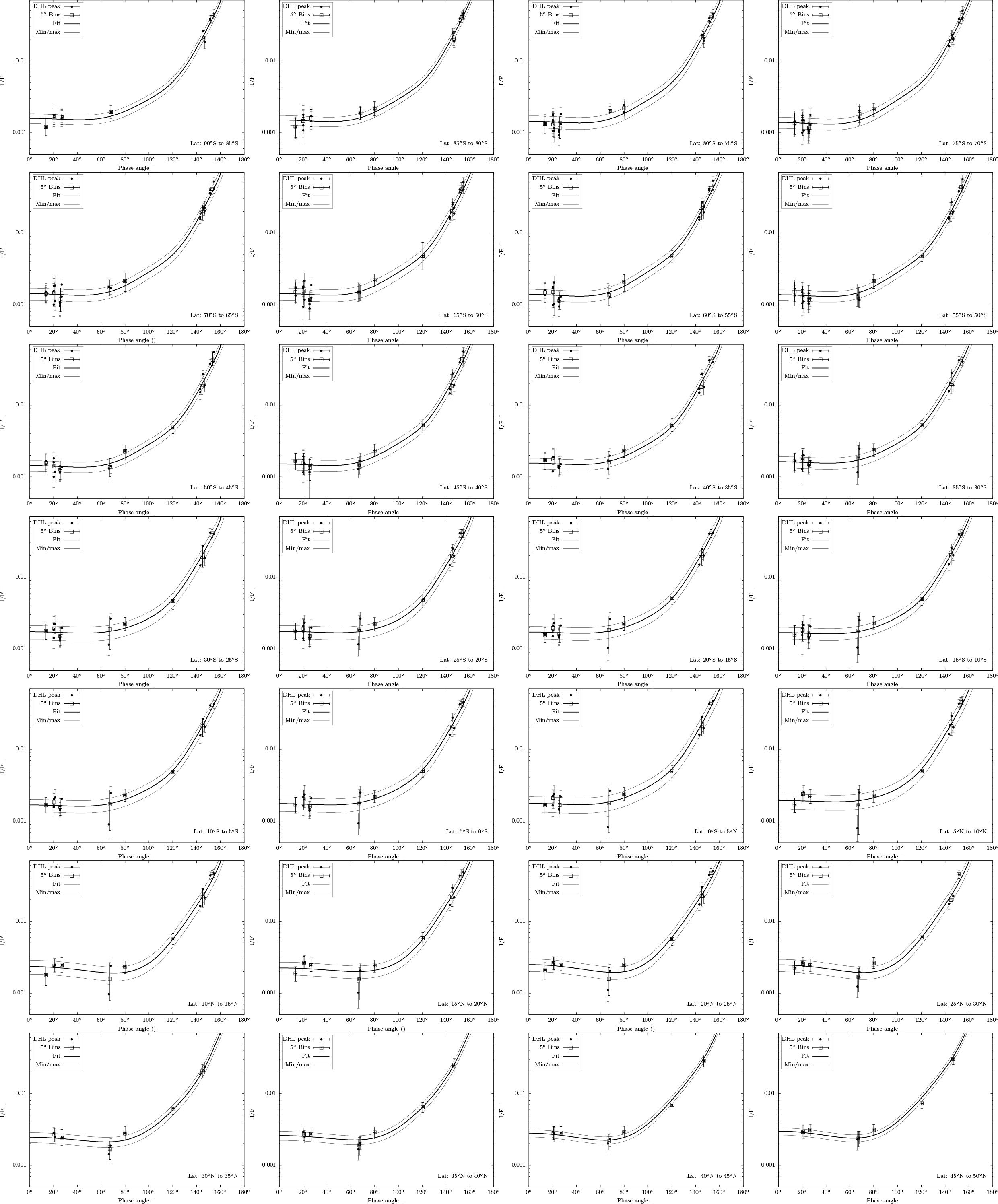}
\caption{\label{fig:all_fits}Best fits of $I/F$ sorted by 5$^\circ$ latitude bins between 90$^\circ$S and 50$^\circ$N. For each panel, the filled circles with associated error bars represent the observed $I/F$, and the black solid line corresponds to the best fit on R$_m$, $N$, and N$_\mathrm{los}$. The two gray solid lines represent the best fits in the extreme cases of minimum and maximum observed $I/F$. We can note that the fit is less consistent on the last two rows ($>10^\circ$N) when the number of observations decreases.}
\end{figure*}

The complete dataset of the best fits for each latitude bin is shown in Figure~\ref{fig:all_fits}. We notice that the fits are very similar for all latitudes lower than 10$^\circ$N. At higher latitudes, the shape of the phase function changes and presents an increase in the backscattered direction (i.e. at low phase angle). This effect is considered to be an artifact due to the lower number of images available at these latitudes.

\begin{figure*}[ht!]\centering
\includegraphics[width=\textwidth]{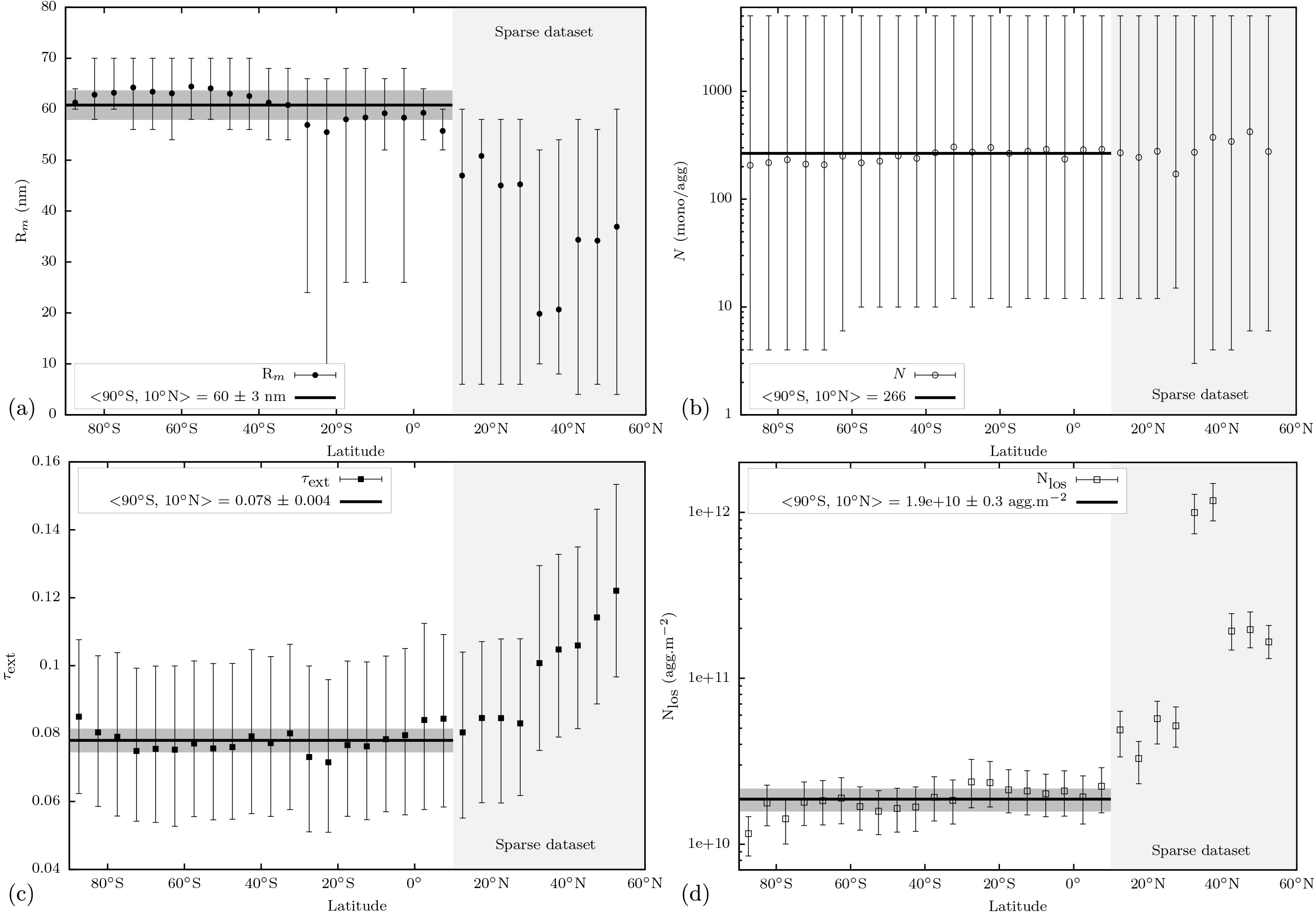}
\caption{\label{fig:res}Latitudinal retrieval of (a) monomer radius, (b) number of monomers per aggregate, (c) tangential opacity, and (d) tangential column number density derived from the $\chi^2$ exploration with a confidence level $\Delta\chi^2 < 1.00$. The circle and squares represents the local mean value (calculated in log-space for $N$ and $N_\mathrm{los}$). The solid lines and the drak gray boxes correspond to the mean and the standard deviation of these mean values between 90$^\circ$S and 10$^\circ$N where the dataset is most constraining. Latitudes in the light gray area are excluded due to their sparse dataset (winter in the northern hemisphere, cf. Tab.~\ref{tab:1}). The upper error-bars on $N$ at 5,000 monomers per aggregate correspond to typical aggregates observed close to the surface \citep{Tomasko2009}.}
\end{figure*}

The parameter values derived for the different latitude bins are summarized in figure~\ref{fig:res}. Each point represents the average value inside the area covered by $\Delta\chi^2<1.00$ and the error bars correspond to the extreme values with the same confidence level. The latitudinal variation of these parameters can be split into two main areas: below and above 10$^\circ$N. Below 10$^\circ$N, the monomer radius and the tangential opacity can be considered constant with R$_m = 60 \pm 3$~nm and $\tau_\mathrm{ext} = 0.078 \pm 0.004$. The number of monomers per aggregate is larger than ten monomers without upper constraint. We use $N = 266$ as a latitude mean (in the log-space). All the latitude means below 10$^\circ$N are summarized in Table~\ref{tab:2}. The optical properties derived from the \cite{Tomasko2008} model with these aggregates are listed in Table~\ref{tab:3}. Above 10$^\circ$N, we find a decrease in the aerosols size associated with an increase of the tangential opacity, i.e. smaller and more numerous particles or smaller particles with a higher extinction. However theses results inside the sparse dataset region need to be taken with caution due to the lower number of observation available in the 2005-2007 period as we mention before. For each latitude, we derive the tangential column number density by dividing the tangential opacity with the extinction cross-section calculated with the \cite{Tomasko2008} model. This value is also constant below 10$^\circ$N with N$_\mathrm{los} = 1.9 \pm 0.3 \cdot 10^{10}$~agg.m$^{-2}$ but provide no consistent information above 10$^\circ$N.

\begin{table}[ht!]
\centering
\caption{\label{tab:2}Latitudinal average below 10$^\circ$N of the aerosol properties in the detached haze layer.}
\scriptsize
\begin{tabular}{p{3.35cm} l l l p{.75cm}}
\hline
Parameter                        & Symbol              & Value    & Std   & Unit    \\
\hline
Monomer radius				           & R$_m$	             & 60	      & 3     & nm      \\
Number of monomer per aggregate	 & $N$	               & 266	    & -     & -       \\
Aggregate bulk radius    	       & R$_v$	      	     & 0.4      & -  & $\mu$m  \\
Tangential opacity		           & $\tau_\mathrm{ext}$ & 0.078    & 0.004 & -       \\
Tangential column number density & N$_\mathrm{los}$    & 1.9      & 0.3   & $\times 10^{10}$ m$^{-2}$ \\
\hline
\end{tabular}
\end{table}

\begin{table}[ht!]
\centering
\caption{\label{tab:3}Aerosol optical properties derived from the table~\ref{tab:2} and \cite{Tomasko2008} model (i.e. $\lambda$=338~nm, R$_m$=60~nm, $N$=266, R$_v=0.4\,\mu$m, D$_f$=2.0).}
\scriptsize
\begin{tabular}{l c c r}
\hline
Parameter                   & Symbol                & Value  & Unit   \\
\hline
Scattering cross-section		& $\sigma_\mathrm{sct}$ & 2.9e-12 & m$^2$  \\
Extinction cross-section		& $\sigma_\mathrm{ext}$ & 4.2e-12 & m$^2$  \\
Absorption cross-section		& $\sigma_\mathrm{abs}$ & 1.3e-12 & m$^2$  \\
Single-scattering albedo	   & $\omega$	            & 0.69    & -      \\
\hline
Phase function: $\theta$ = 0$^\circ$ & - & 185.8  &    -   \\
$\theta$ =	 1$^\circ$		&	   -	       & 178.1  &    -   \\
$\theta$ =	 2$^\circ$		&	   -	       & 157.2  &    -   \\
$\theta$ =	 4$^\circ$		&	   -	       & 98.9   &    -   \\
$\theta$ =	 6$^\circ$		&	   -	       & 52.6   &    -   \\
$\theta$ =	 8$^\circ$		&	   -	       & 28.4   &    -   \\
$\theta$ =	10$^\circ$		&	   -	       & 17.4   &    -   \\
$\theta$ =	15$^\circ$		&	   -	       & 8.0    &    -   \\
$\theta$ =	20$^\circ$		&	   -	       & 4.7    &    -   \\
$\theta$ =	30$^\circ$		&	   -	       & 2.2    &    -   \\
$\theta$ =	40$^\circ$		&	   -	       & 1.1    &    -   \\
$\theta$ =	50$^\circ$		&	   -	       & 0.58   &    -   \\
$\theta$ =	60$^\circ$		&	   -	       & 0.36   &    -   \\
$\theta$ =	80$^\circ$		&	   -	       & 0.21   &    -   \\
$\theta$ = 100$^\circ$		&	   -	       & 0.144  &    -   \\
$\theta$ = 120$^\circ$		&	   -	       & 0.117  &    -   \\
$\theta$ = 140$^\circ$		&	   -	       & 0.112  &    -   \\
$\theta$ = 160$^\circ$		&	   -	       & 0.115  &    -   \\
$\theta$ = 180$^\circ$		&	   -	       & 0.117  &    -   \\
\hline
\end{tabular}
\end{table}

Finally, the slope of the $I/F$ vertical profile (Fig.~\ref{fig:data}a) above the detached haze layer provide a measure of the aerosol scale height. The latitude mean value of this scale height ($H = 35$~km) can be used to estimate the geometric attenuation factor of the incoming optical depth in the detached haze layer ($z=500$~km):
\begin{equation}
\tau^0_\mathrm{ext} \approx \tau_\mathrm{ext} \cdot \frac{ H }{ \sqrt{ 2 \pi \cdot (R_T+z) \cdot H } } = 3 \cdot 10^{-3} \ll 1
\end{equation}
confirming that the incoming flux from the Sun is not significantly attenuated down to the detached haze layer. Moreover, this scale height can also be used as a proxy to derive the aggregate local number density ($n$):
\begin{equation}
n \approx \frac{ \mathrm{N}_\mathrm{los} }{ \sqrt{ 2 \pi \cdot (R_T+z) \cdot H } }
\end{equation}
At 500~km, for latitudes lower than 10$^\circ$N, we estimate the local number density inside the detached haze layer at $n \approx 2 \cdot 10^{-2}$~agg.cm$^{-3}$. Then, assuming that the aerosols material has a density of 1~g.cm$^{−3}$ and a fractal dimension of 2.0, we derive an estimation of the mass flux of $\sim 10^{-14}$~g.cm$^{-2}$.s$^{-1}$ for these aerosols \citep{Lavvas2010}. Above 10$^\circ$N the data are not consistent.

\section{Discussion}
During this analysis, we have decided to keep the fractal dimension (D$_f$) of the aerosols fixed at 2.0 owing to the \cite{Tomasko2008} model. Other models, like~\cite{Rannou1997} provide a semi-empirical model of absorption and scattering by isotropic fractal aggregates of spheres for higher fractal dimension. However, after investigation we notice that the distribution of pairs of monomers \citep[equation A3]{Rannou1997} does not take into account the cumulative number of dimers (i.e. monomer touching each other). When we compare outputs with T-matrix cases provided by \cite{Tomasko2008}, we notice a significant difference in the tail of the phase function in the backscattering direction at short wavelength. Therefore, without any additional constraints from the observations, we decided to discard this model and keep the fractal dimension fixed at 2.0.

We also considered the sensitivity of our retrievals on the wavelengths sampled by the instrument. The CL1-UV3 filter on the ISS Narrow Angle Camera is centered at 338~nm with a bandwidth of 68~nm. We find a difference smaller than 5\% in the $I/F_\mathrm{synt}$ calculation trough the transmission filter compared with the single central wavelength model presented in this study.

\begin{figure*}[ht!]\centering
\includegraphics[width=\textwidth]{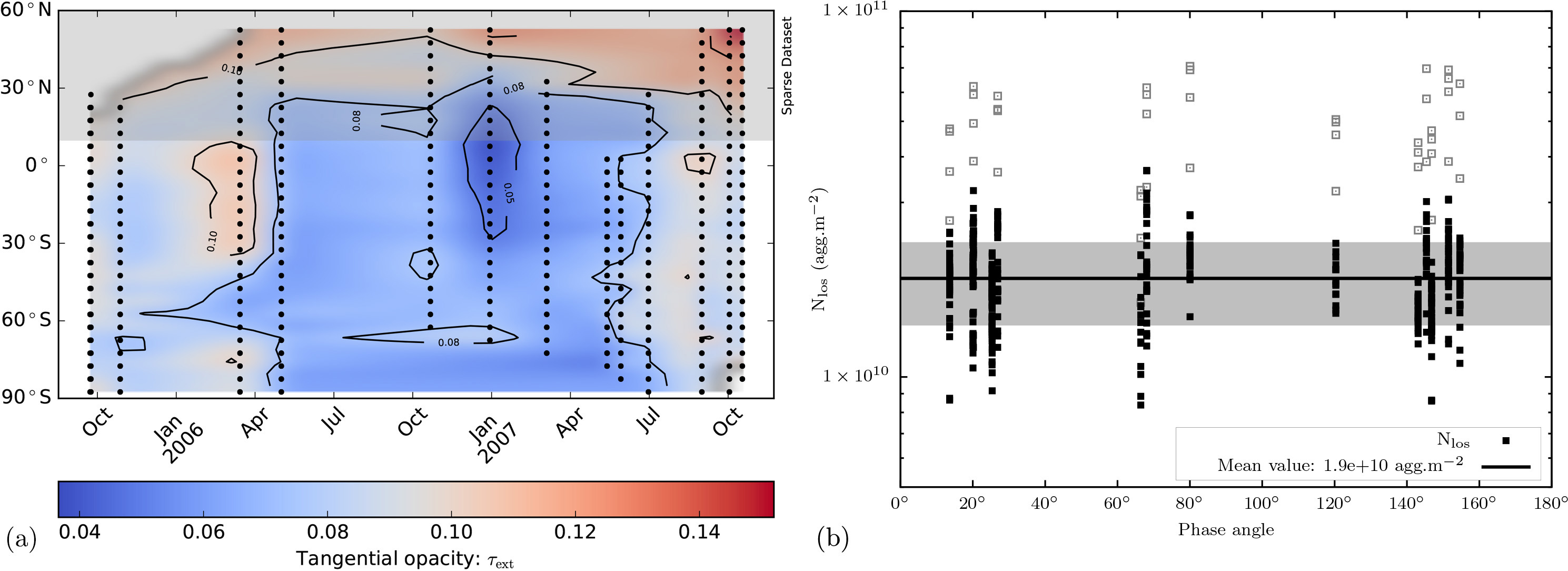}
\caption{\label{fig:time}(a) Temporal/latitudinal map of tangential opacity ($\tau_\mathrm{ext}$) in the detached haze layer. The black dots correspond to the analysis made here and we linearly interpolate for intermittent times. We observe a temporal variability of a factor of $\sim$3 at the equator (see text for detail). We also added the sparse dataset output as a reference, however these latitudes must be taken with caution due to their poor determination of the (R$_m$,$N$) parameters. (b) Tangential column number density (N$_\mathrm{los}$) retrieved for each image as a function of phase angle. The mean over the phase angle and the standard deviation are represented by the solid line and the dark gray box, respectively. As expected, this parameter is independent of the phase angle where the dataset is enough constraining. The open squares represent the values retrieved on the sparse dataset (excluded from the mean).}
\end{figure*}

The 60 $\pm$ 3~nm for the monomer radius of the aerosols retrieved in this analysis is larger than the $40 \pm 10$~nm monomer observed in polarization by DISR \citep{Tomasko2009} in the main haze at lower altitudes ($<$150~km). However, our value is consistent with the 60~nm derived with polarimetry \citep{West1991} and the 66~nm \citep{Rannou1997} retrieved at higher altitude ($>350$~km) from Voyager observations. On the other hand, due to the poor determination of the number of monomers per aggregates, it is not possible to provide any constraints on the size distribution of the aerosols. Narrow and broad log-normal distributions where tested (with a scale parameter $\sigma_0 =$~0.3 and 0.5) but did not provide significantly different results compared to the mono-dispersed case. Moreover, we made a simple comparison with UVIS observations \citep{Koskinen2011}. The comparison between these instruments can help to constrain the aerosol distribution including the small, none scattering aerosols in the detached haze layer \citep{Cours2011}. The extinction tangential opacity retrieved at the local maximum around 500~km is about 0.75 at 190 nm (T41 and T53 flybys in 2008). We extend our results at this wavelength using the \cite{Khare1984} indices and we find a scattering tangential opacity between 0.07 and 0.11. As expected the tangential opacity found with ISS is smaller than the one found with UVIS which include small none scattering aerosols.

The fractal aggregates found in this study are composed of more than ten monomers of 60~nm. As expected from the GCM of Titan this type of aggregates needs long timescales to be formed and will be homogeneously distributed in longitude by the zonal wind \citep{Rannou2004,Larson2015}. As we showed, during the 2005-2007 period the detached haze layer is also very stable with latitude from the south pole up to the 10$^\circ$N with a possible decrease in particle size at higher latitudes where the dataset is less constraining. Finally, we use the optical properties derived from this analysis latitude by latitude to get an estimation of the temporal/longitudinal variability of the tangential opacity of the detached haze layer (Fig.~\ref{fig:time}a). The global variability is low at $0.08 \pm 0.1$ but we notice two patterns around the equator on the images taken in March 2006 and December 2006 (N1521213736\textunderscore 1 and N1546223487\textunderscore 1). The variability on the tangential opacity is about a factor 3 between these two structures (0.04 to 0.12). Both share the same acquisition geometry, a similar phase angle of observation (68 and 66$^\circ$), a same local time (10h28 and 13h28) and a same position on the orbit (202$^\circ$ and 254$^\circ$ to aphelion), only the acquisition time and the longitude (21$^\circ$ and 100$^\circ$W) are significantly distinct between these two images. Theses variations can be attributed to fluctuations of the tangential column number density. For now, its origin is unknown, but our best guess would be the propagation of waves. However, the large gaps between flyby acquisitions do not allow us to derive any temporal constraint. We also checked that our retrievals on the tangential column number density are not biased by the phase angle sampling. When we exclude from the statistics the observations corresponding to the sparser portions of the dataset, then the retrievals have the same spread at low, middle and high phase angle, and do not present any bias (Fig.~\ref{fig:time}b).

\section{Conclusions}
The analysis of Cassini UV images of Titan taken between 2005 and 2007 enable us to verify the stability of the detached haze layer at $500 \pm 8$~km for all latitudes lower than 45$^\circ$N during this period. Comparing the scattered intensity of the detached haze at its maximum for different phase angles with the synthetic intensity based on the \cite{Tomasko2008} model, we were able to constrain the optical properties of aerosols in the detached haze layer. We find that theses aerosols can have at least ten monomers of $60$~nm radius for latitudes lower than 10$^\circ$N. The typical tangential column number density was constrained at $2\cdot 10^{10}$~agg.m$^{-2}$ in the same latitude range. At higher latitudes, we have shown that the size of the aggregates tends to decrease but the lower number of images available in this region due to observation limitations did not allow us to derive better constrains. Then, we used these aggregates (averaged between 90$^\circ$S and 10$^\circ$N) as a homogeneous content to get the local temporal/longitudinal variability of the detached layer during the 2005-2007 period and notice that the tangential opacity can fluctuate up to a factor 3. For now the cause of theses perturbations of the local density are still unknown but they seem to be linked to short scale temporal and/or longitudinal events. Finally, we stress the fact that fractal aerosols can have a fractal dimension higher than 2.0 but the current models are not able to investigate this property without more additional constraints. Observations at different wavelengths and with polarization filters could be used to provide these additional constraints in the future.

\section*{Acknowledgments}

This work was supported by the French ministry of public research. The authors also thank the Programme National de Plan\'{e}tologie (PNP), the Agence National de la Recherche (ANR project "Apostic" No. 11BS65002, France) and the Centre National d'\'{E}tude Spatial (CNES) for their financial support. R.A. West would like to acknowledge the Region Champagne-Ardenne for its support (program "expertise de chercheurs invit\'{e}s" AO 2015).


\bibliography{Biblio}

\begin{thebibliography}{26}
\providecommand{\natexlab}[1]{#1}
\providecommand{\url}[1]{\texttt{#1}}
\providecommand{\href}[2]{#2}
\providecommand{\path}[1]{#1}
\providecommand{\eprint}[1]{\href{http://arxiv.org/abs/#1}{\path{#1}}}
\providecommand{\DOIprefix}{doi:}
\providecommand{\ISBNprefix}{ISBN:}
\providecommand{\ArXivprefix}{arXiv:}
\providecommand{\URLprefix}{URL: }
\providecommand{\Pubmedprefix}{pmid:}
\providecommand{\doi}[1]{\href{http://dx.doi.org/#1}{\path{#1}}}
\providecommand{\isbn}[1]{\href{https://openlibrary.org/search?isbn=#1}{#1}}
\providecommand{\Pubmed}[1]{\href{pmid:#1}{\path{#1}}}
\providecommand{\BIBand}{and}
\providecommand{\bibinfo}[2]{#2}
\ifx\xfnm\undefined \def\xfnm[#1]{\unskip,\space#1}\fi
\bibitem[{Cabane et~al.(1993)Cabane, Rannou, Chassefi{\`{e}}re and
  Israel}]{Cabane1993}
\bibinfo{author}{Cabane M.}, \bibinfo{author}{Rannou P.},
  \bibinfo{author}{Chassefi{\`{e}}re E.}, \bibinfo{author}{Israel G.}
\newblock {\bf \bibinfo{year}{1993}}. \emph{{Fractal aggregates in Titan's
  atmosphere}}.
\newblock \bibinfo{journal}{Planet Sp Sci}
  \bibinfo{volume}{41}(\bibinfo{number}{4}):\bibinfo{pages}{257--267}.
\newblock \DOIprefix\doi{10.1016/0032-0633(93)90021-S}.
\bibitem[{Chassefi{\`{e}}re and Cabane(1995)}]{Chassefiere1995}
\bibinfo{author}{Chassefi{\`{e}}re E.}, \bibinfo{author}{Cabane M.}
\newblock {\bf \bibinfo{year}{1995}}. \emph{{Two formation regions for Titan's
  hazes: indirect clues and possible synthesis mechanisms}}.
\newblock \bibinfo{journal}{Planet Space Sci}
  \bibinfo{volume}{43}(\bibinfo{number}{1-2}):\bibinfo{pages}{91--103}.
\newblock \DOIprefix\doi{10.1016/0032-0633(94)00138-H}.
\bibitem[{Cours et~al.(2011)Cours, Burgalat, Rannou, Rodriguez, Brahic and
  West}]{Cours2011}
\bibinfo{author}{Cours T.}, et~al.
\newblock {\bf \bibinfo{year}{2011}}. \emph{{Dual Origin of Aerosols in Titan's
  Detached Haze Layer}}.
\newblock \bibinfo{journal}{Astrophys J}
  \bibinfo{volume}{741}(\bibinfo{number}{2}):\bibinfo{pages}{L32}.
\newblock \DOIprefix\doi{10.1088/2041-8205/741/2/L32}.
\bibitem[{Griffith et~al.(2006)Griffith, Penteado, Rannou, Brown, Boudon,
  Baines et~al.}]{Griffith2006}
\bibinfo{author}{Griffith C.A.}, et~al.
\newblock {\bf \bibinfo{year}{2006}}. \emph{{Evidence for a Polar Ethane Cloud
  on Titan}}.
\newblock \bibinfo{journal}{Science}
  \bibinfo{volume}{313}(\bibinfo{number}{5793}):\bibinfo{pages}{1620--1622}.
\newblock \DOIprefix\doi{10.1126/science.1128245}.
\bibitem[{Karkoschka(1994)}]{Karkoschka1994}
\bibinfo{author}{Karkoschka E.}
\newblock {\bf \bibinfo{year}{1994}}. \emph{{Spectrophotometry of the Jovian
  Planets and Titan at 300- to 1000-nm Wavelength: The Methane Spectrum}}.
\newblock \bibinfo{journal}{Icarus}
  \bibinfo{volume}{111}(\bibinfo{number}{1}):\bibinfo{pages}{174--192}.
\newblock \DOIprefix\doi{10.1006/icar.1994.1139}.
\bibitem[{Karkoschka(1998)}]{Karkoschka1998}
\bibinfo{author}{Karkoschka E.}
\newblock {\bf \bibinfo{year}{1998}}. \emph{{Methane, Ammonia, and Temperature
  Measurements of the Jovian Planets and Titan from CCD–Spectrophotometry}}.
\newblock \bibinfo{journal}{Icarus}
  \bibinfo{volume}{133}(\bibinfo{number}{1}):\bibinfo{pages}{134--146}.
\newblock \DOIprefix\doi{10.1006/icar.1998.5913}.
\bibitem[{Khare et~al.(1984)Khare, Sagan, Arakawa, Suits, Callcott and
  Williams}]{Khare1984}
\bibinfo{author}{Khare B.N.}, et~al.
\newblock {\bf \bibinfo{year}{1984}}. \emph{{Optical constants of organic
  tholins produced in a simulated Titanian atmosphere - From soft X-ray to
  microwave frequencies}}.
\newblock \bibinfo{journal}{Icarus} \bibinfo{volume}{60}.
\newblock \DOIprefix\doi{10.1016/0019-1035(84)90142-8}.
\bibitem[{Koskinen et~al.(2011)Koskinen, Yelle, Snowden, Lavvas, Sandel,
  Capalbo et~al.}]{Koskinen2011}
\bibinfo{author}{Koskinen T.T.}, et~al.
\newblock {\bf \bibinfo{year}{2011}}. \emph{{The mesosphere and lower
  thermosphere of Titan revealed by Cassini/UVIS stellar occultations}}.
\newblock \bibinfo{journal}{Icarus}
  \bibinfo{volume}{216}(\bibinfo{number}{2}):\bibinfo{pages}{507--534}.
\newblock \DOIprefix\doi{10.1016/j.icarus.2011.09.022}.
\bibitem[{Larson et~al.(2015)Larson, Toon, West and Friedson}]{Larson2015}
\bibinfo{author}{Larson E.J.}, \bibinfo{author}{Toon O.B.},
  \bibinfo{author}{West R.A.}, \bibinfo{author}{Friedson A.J.}
\newblock {\bf \bibinfo{year}{2015}}. \emph{{Microphysical modeling of Titan's
  detached haze layer in a 3D GCM}}.
\newblock \bibinfo{journal}{Icarus}
  \bibinfo{volume}{254}:\bibinfo{pages}{122--134}.
\newblock \DOIprefix\doi{10.1016/j.icarus.2015.03.010}.
\bibitem[{Lavvas et~al.(2009)Lavvas, Yelle and Vuitton}]{Lavvas2009}
\bibinfo{author}{Lavvas P.}, \bibinfo{author}{Yelle R.V.},
  \bibinfo{author}{Vuitton V.}
\newblock {\bf \bibinfo{year}{2009}}. \emph{{The detached haze layer in Titan's
  mesosphere}}.
\newblock \bibinfo{journal}{Icarus}
  \bibinfo{volume}{201}(\bibinfo{number}{2}):\bibinfo{pages}{626--633}.
\newblock \DOIprefix\doi{10.1016/j.icarus.2009.01.004}.
\bibitem[{Lavvas et~al.(2010)Lavvas, Yelle and Griffith}]{Lavvas2010}
\bibinfo{author}{Lavvas P.}, \bibinfo{author}{Yelle R.V.},
  \bibinfo{author}{Griffith C.a.}
\newblock {\bf \bibinfo{year}{2010}}. \emph{{Titan's vertical aerosol structure
  at the Huygens landing site: Constraints on particle size, density, charge,
  and refractive index}}.
\newblock \bibinfo{journal}{Icarus}
  \bibinfo{volume}{210}:\bibinfo{pages}{832--842}.
\newblock \DOIprefix\doi{10.1016/j.icarus.2010.07.025}.
\bibitem[{Lebonnois et~al.(2012)Lebonnois, Burgalat, Rannou and
  Charnay}]{Lebonnois2012}
\bibinfo{author}{Lebonnois S.}, \bibinfo{author}{Burgalat J.},
  \bibinfo{author}{Rannou P.}, \bibinfo{author}{Charnay B.}
\newblock {\bf \bibinfo{year}{2012}}. \emph{{Titan global climate model: A new
  3-dimensional version of the IPSL Titan GCM}}.
\newblock \bibinfo{journal}{Icarus}
  \bibinfo{volume}{218}(\bibinfo{number}{1}):\bibinfo{pages}{707--722}.
\newblock \DOIprefix\doi{10.1016/j.icarus.2011.11.032}.
\bibitem[{Porco et~al.(2005)Porco, Baker, Barbara, Beurle, Brahic, Burns
  et~al.}]{Porco2005}
\bibinfo{author}{Porco C.C.}, et~al.
\newblock {\bf \bibinfo{year}{2005}}. \emph{{Imaging of Titan from the Cassini
  spacecraft.}}
\newblock \bibinfo{journal}{Nature}
  \bibinfo{volume}{434}:\bibinfo{pages}{159--168}.
\newblock \DOIprefix\doi{10.1038/nature03436}.
\bibitem[{Press et~al.(1992)Press, Flannery, Teukolsky and
  Vetterling}]{Press1992}
\bibinfo{author}{Press W.H.}, \bibinfo{author}{Flannery B.P.},
  \bibinfo{author}{Teukolsky S.A.}, \bibinfo{author}{Vetterling W.T.}
\newblock {\bf \bibinfo{year}{1992}}. \bibinfo{title}{{Numerical Recipes in
  Fortran 77: The Art of Scientific Computing}}; vol.~\bibinfo{volume}{1}.
\newblock \bibinfo{edition}{2nd} ed.
\newblock \ISBNprefix\isbn{\bibinfo{isbn}{052143064X}}.
\bibitem[{Rages and Pollack(1983)}]{Rages1983b}
\bibinfo{author}{Rages K.}, \bibinfo{author}{Pollack J.}
\newblock {\bf \bibinfo{year}{1983}}. \emph{{Vertical distribution of
  scattering hazes in Titan's upper atmosphere}}.
\newblock \bibinfo{journal}{Icarus}
  \bibinfo{volume}{55}(\bibinfo{number}{1}):\bibinfo{pages}{50--62}.
\newblock \DOIprefix\doi{10.1016/0019-1035(83)90049-0}.
\bibitem[{Rannou et~al.(1997)Rannou, Cabane, Botet and
  Chassefi{\`{e}}re}]{Rannou1997}
\bibinfo{author}{Rannou P.}, \bibinfo{author}{Cabane M.},
  \bibinfo{author}{Botet R.}, \bibinfo{author}{Chassefi{\`{e}}re E.}
\newblock {\bf \bibinfo{year}{1997}}. \emph{{A new interpretation of scattered
  light measurements at Titan's limb}}.
\newblock \bibinfo{journal}{J Geophys Res Planets}
  \bibinfo{volume}{102}(\bibinfo{number}{E5}):\bibinfo{pages}{10997--11013}.
\newblock \DOIprefix\doi{10.1029/97JE00719}.
\bibitem[{Rannou et~al.(2002)Rannou, Hourdin and McKay}]{Rannou2002}
\bibinfo{author}{Rannou P.}, \bibinfo{author}{Hourdin F.},
  \bibinfo{author}{McKay C.P.}
\newblock {\bf \bibinfo{year}{2002}}. \emph{{A wind origin for Titan's haze
  structure}}.
\newblock \bibinfo{journal}{Nature}
  \bibinfo{volume}{418}(\bibinfo{number}{6900}):\bibinfo{pages}{853--856}.
\newblock \DOIprefix\doi{10.1038/nature00961}.
\bibitem[{Rannou et~al.(2004)Rannou, Hourdin, McKay and Luz}]{Rannou2004}
\bibinfo{author}{Rannou P.}, \bibinfo{author}{Hourdin F.},
  \bibinfo{author}{McKay C.P.}, \bibinfo{author}{Luz D.}
\newblock {\bf \bibinfo{year}{2004}}. \emph{{A coupled dynamics-microphysics
  model of Titan's atmosphere}}.
\newblock \bibinfo{journal}{Icarus}
  \bibinfo{volume}{170}(\bibinfo{number}{2}):\bibinfo{pages}{443--462}.
\newblock \DOIprefix\doi{10.1016/j.icarus.2004.03.007}.
\bibitem[{Smith et~al.(1981)Smith, Soderblom, Beebe, Boyce, Briggs, Bunker
  et~al.}]{Smith1981}
\bibinfo{author}{Smith B.A.}, et~al.
\newblock {\bf \bibinfo{year}{1981}}. \emph{{Encounter with Saturn: Voyager 1
  Imaging Science Results}}.
\newblock \bibinfo{journal}{Science}
  \bibinfo{volume}{212}(\bibinfo{number}{4491}):\bibinfo{pages}{163--191}.
\newblock \DOIprefix\doi{10.1126/science.212.4491.163}.
\bibitem[{Smith et~al.(1982)Smith, Soderblom, Batson, Bridges, Inge, Masursky
  et~al.}]{Smith1982}
\bibinfo{author}{Smith B.A.}, et~al.
\newblock {\bf \bibinfo{year}{1982}}. \emph{{A New Look at the Saturn System:
  The Voyager 2 Images}}.
\newblock \bibinfo{journal}{Science}
  \bibinfo{volume}{215}(\bibinfo{number}{4532}):\bibinfo{pages}{504--537}.
\newblock \DOIprefix\doi{10.1126/science.215.4532.504}.
\bibitem[{Tomasko et~al.(2008)Tomasko, Doose, Engel, Dafoe, West, Lemmon
  et~al.}]{Tomasko2008}
\bibinfo{author}{Tomasko M.G.}, et~al.
\newblock {\bf \bibinfo{year}{2008}}. \emph{{A model of Titan's aerosols based
  on measurements made inside the atmosphere}}.
\newblock \bibinfo{journal}{Planet Sp Sci}
  \bibinfo{volume}{56}(\bibinfo{number}{5}):\bibinfo{pages}{669--707}.
\newblock \DOIprefix\doi{10.1016/j.pss.2007.11.019}.
\bibitem[{Tomasko et~al.(2009)Tomasko, Doose, Dafoe and See}]{Tomasko2009}
\bibinfo{author}{Tomasko M.G.}, \bibinfo{author}{Doose L.R.},
  \bibinfo{author}{Dafoe L.E.}, \bibinfo{author}{See C.}
\newblock {\bf \bibinfo{year}{2009}}. \emph{{Limits on the size of aerosols
  from measurements of linear polarization in Titan's atmosphere}}.
\newblock \bibinfo{journal}{Icarus}
  \bibinfo{volume}{204}(\bibinfo{number}{1}):\bibinfo{pages}{271--283}.
\newblock \DOIprefix\doi{10.1016/j.icarus.2009.05.034}.
\bibitem[{Toon et~al.(1992)Toon, McKay, Griffith and Turco}]{Toon1992}
\bibinfo{author}{Toon O.B.}, \bibinfo{author}{McKay C.P.},
  \bibinfo{author}{Griffith C.A.}, \bibinfo{author}{Turco R.P.}
\newblock {\bf \bibinfo{year}{1992}}. \emph{{A physical model of Titan's
  aerosols}}.
\newblock \bibinfo{journal}{Icarus}
  \bibinfo{volume}{95}(\bibinfo{number}{1}):\bibinfo{pages}{24--53}.
\newblock \DOIprefix\doi{10.1016/0019-1035(92)90188-D}.
\bibitem[{West et~al.(2010)West, Knowles, Birath, Charnoz, {Di Nino}, Hedman
  et~al.}]{West2010}
\bibinfo{author}{West R.}, et~al.
\newblock {\bf \bibinfo{year}{2010}}. \emph{{In-flight calibration of the
  Cassini imaging science sub-system cameras}}.
\newblock \bibinfo{journal}{Planet Sp Sci}
  \bibinfo{volume}{58}(\bibinfo{number}{11}):\bibinfo{pages}{1475--1488}.
\newblock \DOIprefix\doi{10.1016/j.pss.2010.07.006}.
\bibitem[{West and Smith(1991)}]{West1991}
\bibinfo{author}{West R.A.}, \bibinfo{author}{Smith P.H.}
\newblock {\bf \bibinfo{year}{1991}}. \emph{{Evidence for aggregate particles
  in the atmospheres of Titan and Jupiter}}.
\newblock \bibinfo{journal}{Icarus}
  \bibinfo{volume}{90}(\bibinfo{number}{2}):\bibinfo{pages}{330--333}.
\newblock \DOIprefix\doi{10.1016/0019-1035(91)90113-8}.
\bibitem[{West et~al.(2011)West, Balloch, Dumont, Lavvas, Lorenz, Rannou
  et~al.}]{West2011}
\bibinfo{author}{West R.A.}, et~al.
\newblock {\bf \bibinfo{year}{2011}}. \emph{{The evolution of Titan's detached
  haze layer near equinox in 2009}}.
\newblock \bibinfo{journal}{Geophys Res Lett}
  \bibinfo{volume}{38}(\bibinfo{number}{6}).
\newblock \DOIprefix\doi{10.1029/2011GL046843}.

\end{thebibliography}

\end{document}